\pdfoutput=1
\documentclass[%
 reprint,
 amsmath,amssymb,
aps,
]{revtex4-1}

\usepackage{graphicx}
\usepackage{dcolumn}
\usepackage{bm}
\usepackage{tabularx}
\usepackage{adjustbox}
\usepackage{lipsum}
\usepackage{bbm}
\usepackage{algorithm}
\usepackage{algpseudocode}
\usepackage[dvipsnames]{xcolor,colortbl}
\usepackage{amsmath,mathrsfs,lineno,amssymb,amsbsy,subfigure}

\graphicspath{{Figures/}}

\begin{document}


\title{Uncertainty Quantification of Tunable Elastic Metamaterials using Polynomial Chaos}

\author{H. Al Ba'ba'a}
\author{S. Nandi}
\author{T. Singh}
\author{M. Nouh}
 \email{Corresponding author: mnouh@buffalo.edu}
\affiliation{%
 Dept. of Mechanical \& Aerospace Engineering, University at Buffalo (SUNY), Buffalo, NY 14260-4400}


\begin{abstract}
Owing to their periodic and intricate configurations, metamaterials engineered for acoustic and elastic wave control inevitably suffer from manufacturing anomalies and deviate from theoretical dispersion predictions. This work exploits the Polynomial Chaos theory to quantify the magnitude and extent of these deviations and assess their impact on the desired behavior. It is shown that uncertainties stemming from surface roughness, tolerances, and other inconsistencies in a metamaterial's unit cell parameters alter the targetted band gap width, location and the confidence level with which it is guaranteed. The effect of uncertainties are projected from a Bloch-wave dispersion analysis of three distinct phononic and resonant cellular configurations and are further confirmed in the frequency response the finite structures. The analysis concludes with a unique algorithm intended to guide the design of metamaterials in the presence of system uncertainties. 
\end{abstract}

\maketitle

\section{Introduction}

Elastic Metamaterials (EMs) are artificially engineered materials that are commonly realized via a periodic variation in their geometry, mechanical properties or placement of local resonators \cite{Hussein2014, bacquet2018chapter, Nouh2015}. Such EMs exhibit unique shock and vibration mitigation capabilities which can be further tuned and altered with the incorporation of active elements \cite{chen2014piezo, airoldi2011design}. However, the need for ideal periodic architectures introduces several challenges in the fabrication precision of EMs which expectedly suffer from anomalies related to machining tolerances, surface roughness, and other inconsistencies \cite{hussein2007_UQ, he_UQ, wu_UQ}. Consequently, realizing a series of perfectly identical unit cells becomes unrealistic in the presence of inevitable manufacturing uncertainties which eventually lead to undesirable performance variations. Motivated by this, the primary goal of this work is to establish a mathematical framework to quantify model uncertainties in different categories of elastic metamaterials and, perhaps more importantly, the effects of such uncertainties on the targetted performance. The analysis presented here is based on the Polynomial Chaos Theory (PCT). The PCT was first introduced by Wiener \cite{Wiener1938} and has since been demonstrated as a robust methodology to quantify uncertainties in a wide range of applications ranging from static to dynamic systems \cite{singh2010polynomial, kim2013wiener, xiu2002wiener, xiu2003modeling, knio2006uncertainty}. The approach relies on developing surrogate models which can emulate the original stochastic system inexpensively and has also been used to determine statistics (for example mean and variance) accurately.

The emphasis here is placed on band gaps, i.e. extended frequency ranges of forbidden wave propagation; a hallmark feature of phononic, acoustic and elastic metamaterials \cite{Huang2010, xiao2012longitudinal, yu2006a,Pai2014,cummer2016controlling}. Such gaps are theoretically predicted in infinitely long periodic structures and are derived from unit-cell-based models which often fail to capture truncation and boundary effects in finite realizations of such metamaterials; the ramifications of which have been extensively reviewed in literature \cite{albabaa2017PZ, albabaa2017PC, sugino2016mechanism, yousefzadeh2015energy}. As such, the analysis conducted here spans both infinite and finite configurations of three distinct types of elastic metamaterials, namely (1) Periodic structures with periodic variations - reminiscent of traditional phononic crystals, (2) A lattice with a periodic elastic foundation, and (3) A locally resonant metamaterial. Uncertainties are evaluated at the unit cell dispersion level and are later corroborated via full-scale simulations of finite metamaterials with a prescribed cell number. Finally, a set of design guidelines are extracted in the form of an algorithm intended to guarantee the fulfillment of a desired band gap in a given metamaterial.

\vspace{-0.6cm}

\section{Mechanics of Tunable Elastic Metamaterials}
\subsection{Generalized Dispersion Analysis}

\begin{figure}[h!]
     \centering
\includegraphics[]{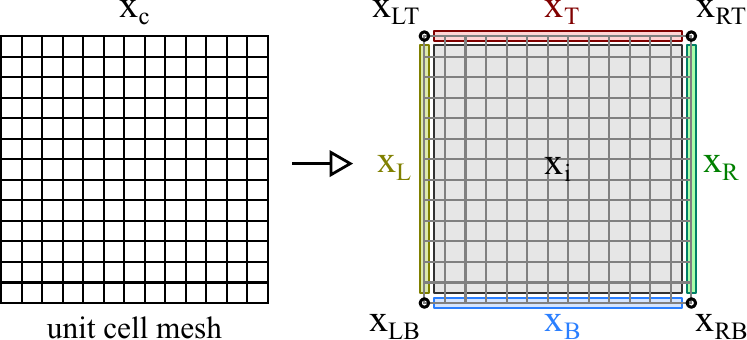}
     \caption{Rearranging the degrees of freedom of a metamaterial unit cell (mesh) based on a finite element model}
     \label{fig:2d_mesh}
\end{figure}

The dynamics of an isolated unit cell of an undamped periodic structure is governed by \begin{equation}
\mathbf{M}_c \ddot{\mathbf{x}}_c(t) + \mathbf{K}_c\mathbf{x}_c(t) = \mathbf{f}_c(t)
\label{eq:UC_EOM}
\end{equation}
where $\mathbf{M}_c$ and $\mathbf{K}_c$ are the mass and stiffness matrices, respectively, whose form depends on the EM type. $(\ \ddot{} \ ) = \frac{d^2}{d t^2}$ denotes the second derivative in time. The formulation used in Eq.~(\ref{eq:UC_EOM}) is well-suited for a finite element implementation to capture the dynamics of the metamaterial's unit cell. Owing to the periodicity of the system, $\mathbf{M}_c$ and $\mathbf{K}_c$ can be condensed by applying Bloch-wave boundary conditions which express the nodal displacements at the unit cell boundaries as an exponential function of the dimensionless wavenumber vector $\bar{\boldsymbol{\beta}}$. For two-dimensional elastic wave propagation (i.e. $\bar{\boldsymbol{\beta}}^{\text{T}} = \{ \bar{\beta}_x \ \bar{\beta}_y \}$), the reduction process requires that the degrees of freedom, given by the nodal deflection vector $\mathbf{x}_c$, be divided into the nine marked on Figure~\ref{fig:2d_mesh} such that
\begin{equation}
\mathbf{x}_c^\text{T} = \{\mathbf{x}_\text{i} \
\mathbf{x}_\text{R} \ \mathbf{x}_\text{L} \
\mathbf{x}_\text{T} \ \mathbf{x}_\text{B} \ 
\mathbf{x}_\text{RT} \
\mathbf{x}_\text{LT} \
\mathbf{x}_\text{RB} \
\mathbf{x}_\text{LB} \}
\label{eq:gen_dof_2d}
\end{equation}

The variables $\bar{\beta}_x$ and $\bar{\beta}_y$ are the nondimensional wavenumbers (spatial frequencies) in the $x$ and $y$ directions and the subscripts i, B, T, L, and R denote the internal, bottom, top, left, and right nodes, respectively. Assuming a harmonic solution and applying the transformation $\mathbf{x}_c = \mathbf{Q} \mathbf{\hat{x}}_c$, the reduced system is computed via post- and pre-multiplying Eq.~(\ref{eq:UC_EOM}) by $\mathbf{Q}$ and its Hermitian transpose $\mathbf{Q}^{\text{H}}$, respectively, to obtain
\begin{equation}
\underbrace{\big [\mathbf{\hat{K}}_c(\bar{\beta}_x,\bar{\beta}_y) - \omega^2 \mathbf{\hat{M}}_c(\bar{\beta}_x,\bar{\beta}_y) \big ]}_{\mathbf{D}_c(\bar{\beta}_x,\bar{\beta}_y,\omega)}
 \mathbf{\hat{x}}_c = \mathbf{0}
\label{eq:disp_eq}
\end{equation}
\noindent where $\mathbf{\hat{x}}_c^\text{T} = \{\mathbf{x}_\text{i} \ \mathbf{x}_\text{R} \
\mathbf{x}_\text{T} \ \mathbf{x}_\text{RT} \}$, $\mathbf{\hat{K}}_c(\bar{\beta}_x,\bar{\beta}_y) = \mathbf{Q}^{\text{H}} \mathbf{K_c} \mathbf{Q}$, $\mathbf{\hat{M}}_c(\bar{\beta}_x,\bar{\beta}_y) = \mathbf{Q}^{\text{H}} \mathbf{M_c} \mathbf{Q}$, and $\mathbf{Q}$ is a transformation matrix given by
\begin{align}
\mathbf{Q}(\bar{\beta}_x,\bar{\beta}_y) = 
\left [ 
\begin{array}{llll}
\mathbf{I}_\text{i} & \mathbf{0} & \mathbf{0} & \mathbf{0} \\ 
\mathbf{0} & \mathbf{I}_\text{R} & \mathbf{0} & \mathbf{0} \\ 
\mathbf{0} & \mathbf{I}_\text{R}e^{-\mathbbm{i}\bar{\beta}_x} & \mathbf{0} & \mathbf{0} \\ 
\mathbf{0} & \mathbf{0} & \mathbf{I}_\text{T} & \mathbf{0} \\ 
\mathbf{0} & \mathbf{0} & \mathbf{I}_\text{T}e^{-\mathbbm{i} \bar{\beta}_y} & \mathbf{0} \\ 
\mathbf{0} & \mathbf{0} & \mathbf{0} & \mathbf{I}_\text{RT}\\ 
\mathbf{0} & \mathbf{0} & \mathbf{0} & \mathbf{I}_\text{RT}e^{-\mathbbm{i} \bar{\beta}_x} \\ 
\mathbf{0} & \mathbf{0} & \mathbf{0} & \mathbf{I}_\text{RT}e^{-\mathbbm{i} \bar{\beta}_y} \\ 
\mathbf{0} & \mathbf{0} & \mathbf{0} & \mathbf{I}_\text{RT}e^{-\mathbbm{i} (\bar{\beta}_x+\bar{\beta}_y)} \\ 
\end{array}
\right ]
\label{eq:Q_transform}
\end{align}
The size of the identity matrices $\mathbf{I}_{( \cdot )}$ in Eq.~(\ref{eq:Q_transform}) depends on the degrees of freedom in $\mathbf{\hat{x}}_c^\text{T}$ and are labeled accordingly. It is noteworthy that the forcing term in Eq.~(\ref{eq:UC_EOM}) vanishes after the Bloch boundary condition is applied since $\mathbf{Q}^{\text{H}} \mathbf{f}_c = \mathbf{0}$ corresponds to an equilibrium state between unit cells \cite{farzbod2011analysis}. Once Eq.~(\ref{eq:disp_eq}) is obtained, the dispersion relation can then be attained by sweeping the values of the wavenumbers across the Irreducable Brillouin Zone (IBZ) and solving the eigenvalue problem in Eq.~(\ref{eq:disp_eq}) for the angular frequency $\omega$. The solution of the eigenvalue problem against each set of wavenumber values yield the metamaterial's dispersion diagram. 

\subsection{Elastic Metamaterial Design}

 \begin{figure*}[]
     \centering
\includegraphics[]{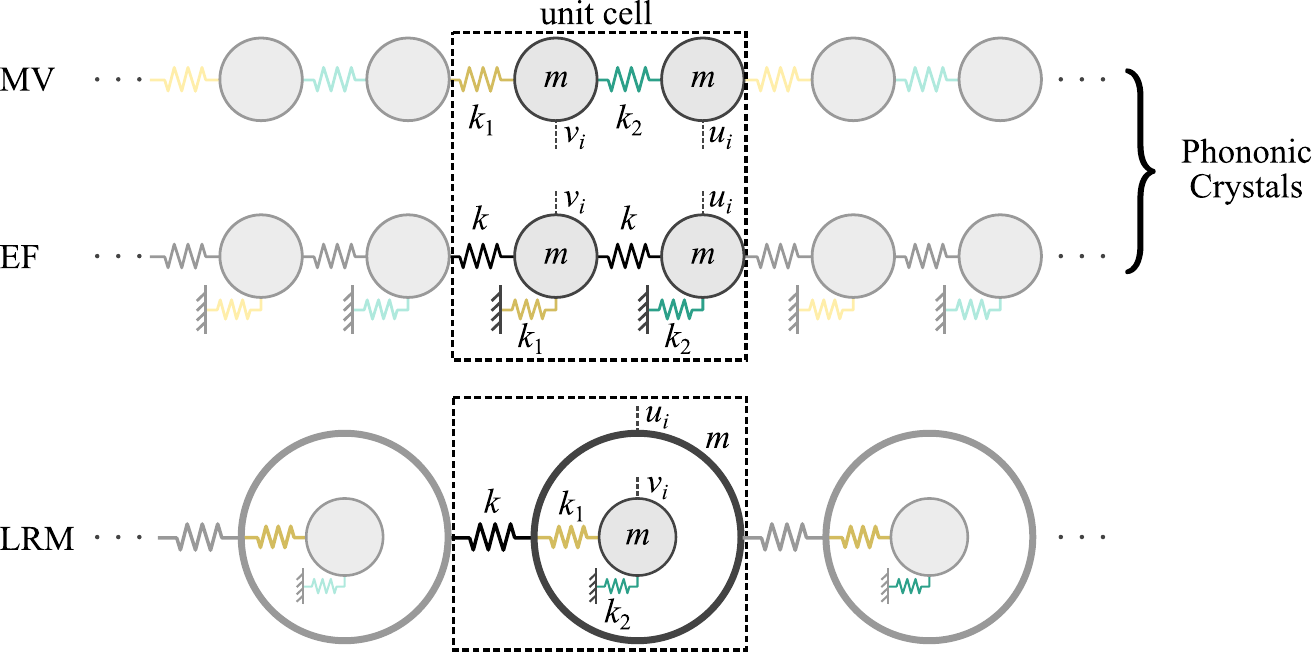}
     \caption{Configuration and design parameters of the three Elastic Metamaterials (EMs) considered}
     \label{fig:sch}
 \end{figure*}
 
The presence of uncertainties in an EM's system parameters hinders its ability to fulfil the desired performance. In here, we focus on the effect of such uncertainties on the EM's dispersion patterns (band structure), dynamic response and its ability to sustain Bragg and local resonance band gaps \cite{liu2012wave, albabaa2017PZ}. For a conceptual understanding, we consider three types of one-dimensional (1D) EMs: (1) A chain of identical masses connected via springs of alternating stiffnesses, (2) A monatomic lattice with a periodicity in the form of alternating grounded springs, and (3) A locally resonant metamaterial (LRM) with grounded springs. Design parameters of all three configurations are shown in Figure~\ref{fig:sch}. The first two types represent different configurations of phononic crystals and, for simplicity, are henceforth referred to as MV and EF in reference to their material variation and elastic foundation, respectively. For all three cases, i.e. MV, EF and LRM, we adopt the following parameterization for the spring constants: $k_{1} = k(1+\vartheta)$ and $k_{2} = k(1-\vartheta)$, where $\vartheta$ can be regarded as a control parameter that resembles an active element in EMs (e.g. piezoelectric material \cite{chen2014piezo,nouh2016periodic}) while $k = k_0$ represents a passive stiffness constant. This parameterization of stiffness coefficients has also been exploited in the design of topologically protected lattices \cite{pal2018amplitude}. Eqs.~(\ref{eq:UC_EOM}) through (\ref{eq:Q_transform}) for the dispersion relation remain intact with the definition of the displacement and forcing vectors being $\mathbf{x}_c^\text{T} = \{\mathbf{x}_\text{i} \ 
\mathbf{x}_\text{R} \ \mathbf{x}_\text{L} \} = \{v_{i} \ u_{i} \ u_{i-1} \}$ and $\mathbf{f}_c^\text{T}= \{ 0 \ f_{i} \ f_{i-1}\}$. As a result, the transformation matrix $\mathbf{Q}$ reduces to
\begin{align}
\mathbf{Q}=
\begin{bmatrix}
1 & 0\\
0 & 1 \\
0 & e^{-\mathbbm{i} \bar{\beta}_x}\\
\end{bmatrix} 
\end{align}
and, consequently, the condensed nodal displacement vector is $\hat{\mathbf{x}}_c^\text{T} = \{\mathbf{x}_\text{i} \ \mathbf{x}_\text{R} \} = \{v_{i} \ u_{i} \}$. The mass and stiffness matrices corresponding to the three EM types are provided in Appendix~\ref{app:M_and_K}. Each of the considered unit cells comprises two degrees of freedom and, as such, the determinant of $\mathbf{D}_c(\bar{\beta}_x,\omega)$ derived from Eq.~(\ref{eq:disp_eq}) yields a biquadratic equation which takes the following form
\begin{equation}
    \alpha_2 \omega^4 - \alpha_1 \omega^2 + \alpha_0 = 0
    \label{eq:disp_gen_form}
\end{equation}

Introducing $\omega_0=\sqrt{k_0/m}$ and implementing the parameterization presented earlier, the coefficient $\alpha_2$ becomes unity and the rest of the dispersion relation constants, i.e. $\alpha_1$ and $\alpha_0$ are listed in Table~\ref{Tb:Dispersions} for all three cases. The roots of Eq.~(\ref{eq:disp_gen_form}) generate two dispersion branches and their analytical expressions can be written in the following form (for brevity, we drop the subscript $x$ from $\bar{\beta}_x$ throughout the discussion of 1D systems)
\begin{equation}
    \omega_\pm(\vartheta,\bar{\beta}) = \omega_0 \Omega_\pm (\vartheta,\bar{\beta})
    \label{eq:gen_branches}
\end{equation}
where $\Omega_\pm(\vartheta,\bar{\beta})$ is given by
\begin{subequations}
\begin{equation}
   \Omega_\pm (\vartheta,\bar{\beta}) =  \sqrt{2 \pm \sqrt{2+2\vartheta^2+2(1-\vartheta^2) \cos \bar{\beta}}}
\end{equation}
\begin{equation}
   \Omega_\pm (\vartheta,\bar{\beta}) =  \sqrt{3 \pm \sqrt{2+\vartheta^2+2 \cos \bar{\beta}}}
\end{equation}
\begin{widetext}
\begin{equation}
   \Omega_\pm (\vartheta,\bar{\beta}) = \frac{1}{\sqrt{2}}   \sqrt{3+\vartheta+4\sin^2 \frac{\bar{\beta}}{2} \pm \sqrt{(5+6\vartheta+5\vartheta^2)+
   8(\vartheta-1)\sin^2\frac{\bar{\beta}}{2} + 16 \sin^4\frac{\bar{\beta}}{2}}}
\end{equation}
\end{widetext}
\label{eq:PM_branches}
\end{subequations}
for MV, EF and LRM, respectively. Of specific interest here are the limits of the emerging band gaps with respect to the control parameter $\vartheta$, which are conveniently derived from the dispersion branches, i.e. Eq.~(\ref{eq:PM_branches}), and summarized in Table~\ref{Tb:BG_Limits}. In all three cases, a band gap of width $\Delta \Omega = \Omega_u - \Omega_l$ splits the acoustical and optical branches of the dispersion relation, where $\Omega_u$ and $\Omega_l$ denote the upper and lower limits, respectively, normalized by $\omega_0$. An additional band gap opens up in the EF and LRM cases as a result of the grounded elastic supports \cite{albabaa2017PC}, which starts at $\Omega = 0$ and ends at $\Omega_z$.

\begin{table}[h!]
\centering
\caption{Dispersion relation coefficients $\alpha_0$ and $\alpha_1$}
\begin{tabular}{l @{\hspace{7.0ex}} l @{\hspace{7.0ex}} l}
\hline\hline\
& $\alpha_1$ & $\alpha_0$ \\
\hline 
MV & $4 \omega_0^2$ & $ 4\omega_0^4(1-\vartheta^2) \sin^2 \frac{\bar{\beta}}{2}$ \\
\hline 
EF & $6 \omega_0^2$& $\omega_0^4 \left(7 - \vartheta^2 - 2\cos \bar{\beta} \right)$ \\
\hline
LRM & $(3+\vartheta+4\sin^2 \frac{\bar{\beta}}{2})$ & $\omega_0^4 \left(1-\vartheta^2+2\sin^2 \frac{\bar{\beta}}{2}\right)$ \\
\hline\hline
\label{Tb:Dispersions}
\end{tabular}
\end{table}

\begin{figure*}[]
     \centering
\includegraphics[]{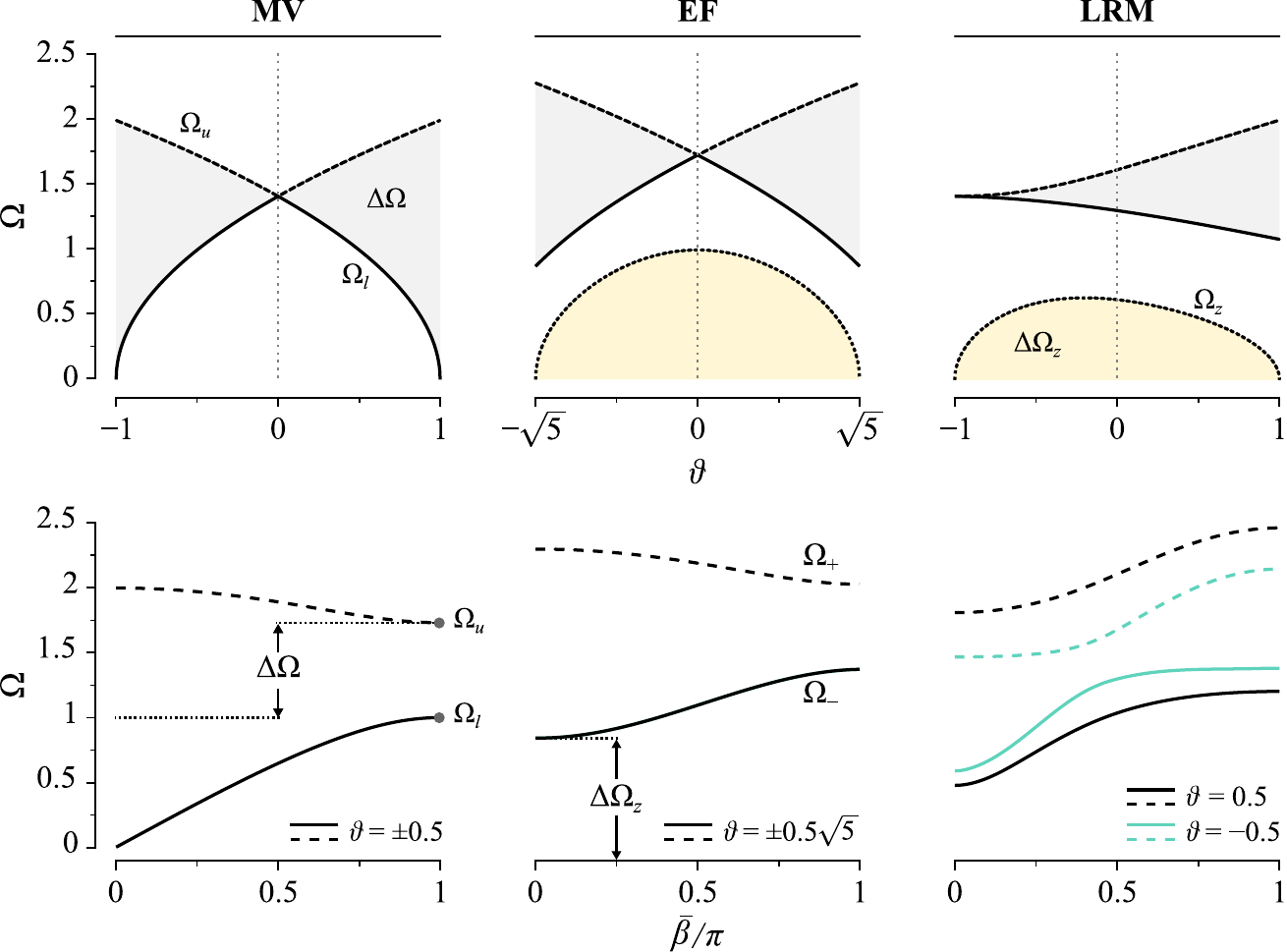}
     \caption{\textit{Top:} Band gap limits as a function of the control parameter $\vartheta$ for the three EMs depicted in Figure~\ref{fig:sch}. \textit{Bottom:} Examples of the dispersion diagrams at $\vartheta/{\vartheta}_{\max} = \pm 0.5$ (Note that $\vartheta_{\max}=1$ for the MV and LRM cases, and $\sqrt{5}$ for the EF)}
     \label{fig:Theta_BG}
 \end{figure*}

The top row of Figure~\ref{fig:Theta_BG} depicts the variation of the aforementioned band gap limits for a range of $\vartheta$ values, with $\Omega = \frac{\omega}{\omega_0}$. The first observation is that the range within which the control parameter $\vartheta$ is allowed to vary is confined and differs from one metamaterial configuration to another. Both the MV and LRM respond stably to a stimulation in the range $\vartheta \in [-1,1]$, while the EF is capable of going up to $\vartheta \in [-\sqrt{5},\sqrt{5}]$ without stability issues. Exceeding the allowable value $\vartheta_\text{max}$ in each case renders the system unstable as a consequence of $\mathbf{K}_c$ becoming negative definite or negative semi-definite. The second observation is that both the phononic crystal configurations (MV and EF) exhibit a symmetric profile of the band gap limits around $\vartheta = 0$. Consequently, band gap widths are identical for $\pm \vartheta$ in both of these configurations. The LRM, on the other hand, has its first and second band gap widths reach their zenith at nonzero values of $\vartheta=-0.2$ and $1$, respectively. It's also worth noting that the zero frequency band gap in the EF case is maximized when the system is not stimulated (i.e. $\vartheta = 0$) and closes at $\vartheta_\text{max}$. This is in sharp contrast to the second band gap which widens with higher values of $\vartheta$ and reaches its peak at $\vartheta_\text{max}$. Finally, representative examples of the dispersion diagrams (band structures) evaluated at $\vartheta/\vartheta_\text{max} = \pm 0.5$ are provided in the bottom panel of Figure~\ref{fig:Theta_BG}.
\begin{table}[h!]
\centering
\caption{Closed-form expressions for the band gap limits}
\begin{tabular}{l @{\hspace{3.0ex}} l @{\hspace{3.0ex}} l}
\hline\hline\
 & Band Gap & Limits \\ 
 & No. & \\
\hline 
MV & 1 & $ \Omega_{u,l} = \sqrt{2 \pm 2 |\vartheta|}$\\
\hline 
EF & 1 & $\Omega_z = \sqrt{3 - \sqrt{4+\vartheta^2}}$ \\
& 2 & $\Omega_{u,l}=\sqrt{3 \pm |\vartheta|}$ \\
\hline
LRM & 1 & $\Omega_z=\frac{1}{\sqrt{2}} 
    \sqrt{(3+\vartheta) - \sqrt{5+6\vartheta+5\vartheta^2}}$ \\
& 2 & $\Omega_l=\frac{1}{\sqrt{2}} 
    \sqrt{(3+\vartheta) + \sqrt{5+6\vartheta+5\vartheta^2}}$,\\
    && $\Omega_u=\frac{1}{\sqrt{2}} 
    \sqrt{(7+\vartheta) - \sqrt{13+14\vartheta+5\vartheta^2}}$ \\
\hline\hline
\label{Tb:BG_Limits}
\end{tabular}
\end{table}

\subsection{Transfer Functions of Finite EMs}

The framework presented thus far describes the wave propagation aspects of a self-repeating unit cell in an infinite periodic chain; a typical modeling approach in the metamaterials domain. The latter, however, neglects truncation effects, boundary conditions, and the size of finite metamaterial realizations. To capture such effects, we consider the dynamics of an EM with a finite number of cells $n$ and assess the transmission characteristics of the EM via an end-to-end transfer function $\frac{u_n(s)}{f(s)} = K \frac{Z(s)}{P(s)}$, where $K$, $Z(s)$ and $P(s)$ represent the gain, zeros polynomial and poles polynomial, respectively. We start by stating the general motion equations of a finite EM in compact matrix form
\begin{equation}
\mathbf{M} \ddot{\mathbf{x}}(t) + \mathbf{K}\mathbf{x}(t) = \mathbf{f}(t)
\label{eq:FS_EOM}
\end{equation}
where $\mathbf{M} = m \mathbf{I}_{2n}$ is the global mass matrix and $\mathbf{I}_{2n}$ is an identity matrix of size $2n$. The entries of the global stiffness matrix $\mathbf{K}$ depend on the EM type as detailed in Appendix~\ref{App:K_matrices} for a free-free boundary conditions. The displacement vector takes the form $\mathrm{\mathbf{x}}^{\text{T}} = \begin{Bmatrix} v_1 & u_1 & \dots &  v_n &  u_n \end{Bmatrix}$ for the phononic crystal configurations (i.e. MV and EF), and $\mathrm{\mathbf{x}}^{\text{T}} = \begin{Bmatrix} u_1 & \dots & u_n & v_1 & \dots & v_n \end{Bmatrix}$ for the LRM. The forcing vector $\mathrm{\mathbf{f}}^{\text{T}} =\begin{Bmatrix} f(t) & 0 & \dots & 0
\end{Bmatrix}$ captures an excitation applied at on one end of the finite EM. Applying the Laplace transform reduces Eq.~(\ref{eq:FS_EOM}) to $\mathbf{D}(s)\mathbf{x}(s) = \mathbf{f}(s)$ where $\mathbf{D}(s) = \mathbf{M}s^2 + \mathbf{K}$ is the \textit{dynamic} stiffness matrix. The poles of the system can then be found from the determinant of $\mathbf{D}(s)$ and are indifferent to the excitation and measurement (sensing) locations. Analytical expressions for the aforementioned transfer functions for all three EM types considered here and free-free boundary conditions have been recently reported and the derivations are therefore omitted here for brevity \cite{albabaa2017PC,albabaa2017PZ}. However, final expressions for $Z(s)$ and $P(s)$ for the MV, EF, and LRM cases are summarized in Table~\ref{Tb:P(s)_exps}.

\begin{figure*}[]
     \centering
\includegraphics[width=0.95\textwidth]{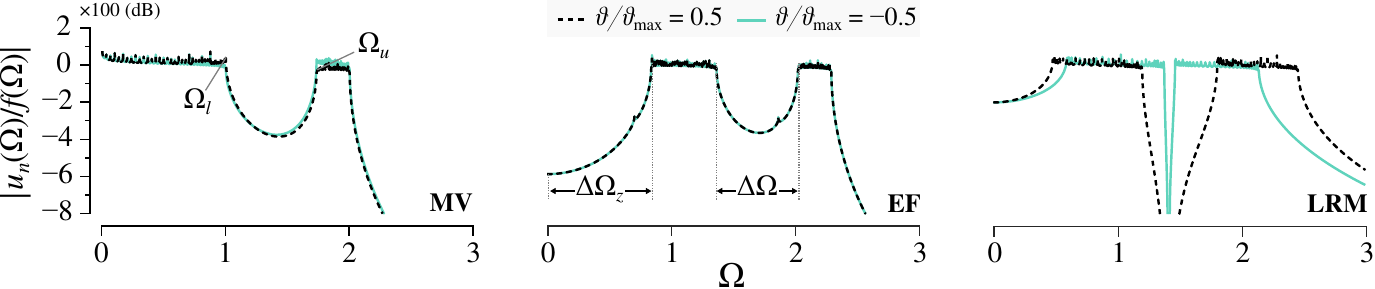}
     \caption{Frequency response of finite realizations of each of the three EMs ($n=40$ cells) at $\vartheta=0.5$ (dashed) and $-0.5$ (solid)}
     \label{fig:FRF_nominal}
 \end{figure*}
 \begin{table*}[t]
\centering
\caption{Analytical expressions for the zeros and poles polynomials, $Z(s)$ and $P(s)$, for the different EM configurations. Note that $\hat{\beta}_i = 4 \sin^2\left(\frac{i-1}{2n}\pi\right)$}
\begin{tabular}{l @{\hspace{10.0ex}} l}
\hline\hline
 & $Z(s)$ \\
\hline 
MV & $(1-\vartheta)(1-\vartheta^2)^{n-1}k^{2n-1}$ (no zeros)\\
EF & $k^{2n-1}$ (no zeros)\\
LRM & $k^{n-1}(ms^2+2k)^n$ \\
\hline
 & \\
 & $P(s)$ \\
\hline 
MV & $ s^2\left(m^2s^2 + 2mk(1-\vartheta)\right)\prod_{i=2}^{n} m^2s^4 + 4mks^2 + k^2(1-\vartheta^2)\hat{\beta}_i$ \\
EF & $(m^2s^4 + 4mks^2 + k^2(3-\vartheta^2))\prod_{i=2}^{n} m^2s^4 + 6mks^2 + k^2(5-\vartheta^2+\hat{\beta}_i)$ \\
LRM & $\prod_{i=1}^n m^2 s^4 + mk(3+\vartheta+\hat{\beta}_i)s^2 + k^2 \left((1-\vartheta^2)+2\hat{\beta}_i\right)$ \\
\hline\hline
\label{Tb:P(s)_exps}
\end{tabular}
\end{table*}
 
The frequency response functions (FRFs) of the three EMs are graphically presented in Figure~\ref{fig:FRF_nominal} for $\vartheta = \pm 0.5 \vartheta_\text{max}$ and $n=40$ cells. The attenuation regions in the FRFs are in excellent agreement with the band gaps predicted by the dispersion relations (Figure~\ref{fig:Theta_BG}). In the finite regime, the band gaps are confined by poles (i.e. natural frequencies) at both ends which also agrees the theoretical limits listed in Table~\ref{Tb:BG_Limits} with sufficiently large $n$ \cite{albabaa2017PC}. Unlike the LRM case, the FRF slightly changes in the MV case and remains unchanged in the EF one as $\vartheta$ flips its sign, owing to the symmetrical response of the band gap limits about $\vartheta=0$ (Figure~\ref{fig:Theta_BG}). Although the dispersion relation of the EF suggests that such system may stay stable up to $|\vartheta| \leq \sqrt{5}$, the response of the finite free-free EF metamaterial will not be stable for any $|\vartheta| > \sqrt{3}$. This is evident from the first term of $P(s)$ in Table~\ref{Tb:P(s)_exps}, where $k^2(3-\vartheta^2)$ becomes negative for $|\vartheta| > \sqrt{3}$ resulting in a pole in the right half $s$-plane.

\section{Uncertainty Analysis Using Polynomial Chaos}
\label{sec:uncertain_sec}
\subsection{Overview}

Following the previous analysis of both the infinite (unit-cell-based) and finite EMs, we now aim to quantify arising uncertainties by utilizing a Polynomial Chaos approach \cite{Wiener1938}. To briefly illustrate the notion of using Polynomial Chaos to represent an uncertain system, consider a stochastic system of the form $Y = f(\eta)$, where $\eta$ is a random input with a known probability density function (PDF) $g(\eta)$ and $f(\eta)$ is an arbitrary non-linear function. The output $Y$ can be approximated by a simple polynomial surrogate model $\tilde{Y}$ with a finite order $n_p$ in terms of the random input $\eta$, such that
\begin{equation}
    Y \approx \tilde{Y} = Y_0\mathcal{P}_0(\eta) + Y_1\mathcal{P}_1(\eta) + \hdots + Y_{n_p}\mathcal{P}_{n_p}(\eta)
    \label{eq:y_tilde}
\end{equation}
where $Y_p$ ($p = 0, 1, \dots, n_p$) are the corresponding coefficients and $\mathcal{P}_p$ are basis functions comprising orthogonal polynomials (which are dependent on the PDF of $\eta$). As the present analysis focuses on a uniformly distributed probability of the random variable, the orthogonal basis functions are chosen as the Legendre polynomials
\begin{equation}
    \mathcal{P}_p(\eta) = \frac{1}{2^p}\sum_{q = 0}^{\lfloor \frac{p}{2} \rfloor} (-1)^q {p \choose q} {2(p-q) \choose p} \eta^{p-2q}
   \label{eq:Leg_poly}
\end{equation}
where $\lfloor \cdot \rfloor$ is the floor function and $\eta \in [-1,1]$. The coefficients $Y_p$ of the surrogate model $\tilde{Y}$ are derived based on the projection of the model error onto the basis function space using the Galerkin projection methodology \cite{kim2013wiener}. This optimization problem leads to the coefficients $Y_p$ and are expressed as the ratio of two integrals
\begin{equation}
    Y_p = \frac{\int_{\mathcal{S}}f(\eta)\mathcal{P}_p(\eta) g(\eta)d\eta}{\int_{\mathcal{S}}\mathcal{P}_p^2(\eta) g(\eta)d\eta}
    \label{eq:y_q}
\end{equation}
where $\mathcal{S}$ denotes the domain of uncertainty. For higher dimension of uncertainty, Eq.~(\ref{eq:y_q}) can be computationally expensive to evaluate. For linear systems, a sampling-based approach which exactly reproduces the coefficients of the surrogate model without having to evaluate multidimensional integrals has been recently developed \cite{nandi_18_nonintrusive}. This can serve as an non-intrusive approach to developing the Polynomial Chaos based surrogate models. 

It is often necessary in various applications to determine the bounds of variation of the uncertain function given a specific finite domain of the input space. The range-enclosing property of Bernstein polynomials is ideal to determine conservative bounds on the range of the output of the function (i.e. $Y$); a property that is not readily available from the Legendre polynomials. Therefore, we seek a unique and linear transformation between two equal order sets of polynomials, i.e. from a $p^{\text{th}}$ Legendre bases to a $p^{\text{th}}$ Bernstein bases expansions. This transformation yields the Bernstein coefficients $Y_p^b$, which satisfies
\begin{equation}
    \tilde{Y} = \sum_{p = 0}^{n_p} Y_p\mathcal{P}_p(\eta) = \sum_{p = 0}^{n_p} Y_p^b\mathcal{B}^{n_p}_p(\eta)
\end{equation}
where the general form of the Bernstein bases is
\begin{equation}
    \mathcal{B}^p_q(\eta) = 
    \begin{pmatrix}
    p \\
    q
    \end{pmatrix}
    \eta^{q}(1-\eta)^{p-q}
\end{equation}

It is worth noting that the input $\eta$, unlike the Legendre polynomials, is limited to the range $\eta \in [0,1]$. Here, we denote the linear transformation between the two bases as $\mathcal{Q}$ such that $\boldsymbol{y}^b = \mathcal{Q}\boldsymbol{y}$, where $\boldsymbol{y}^{b} = \begin{Bmatrix} Y_0^b & Y_1^b  & \dots & Y_{n_p}^b \end{Bmatrix}^{\text{T}}$ and $\boldsymbol{y} = \begin{Bmatrix} Y_0 & Y_1 & \dots & Y_{n_p} \end{Bmatrix}^{\text{T}}$. An analytical form of the entries of the transformation $\mathcal{Q}$ are given by \cite{farouki2000legendre}
\begin{equation}
\mathcal{Q}_{ql} = \frac{1}{ {p \choose l}} \sum_{\hat{q} = \max(0,q+l-p)}^{\min(q,l)} (-1)^{\hat{q}+l} {q \choose \hat{q}} {l \choose \hat{q}} {p-l \choose q-\hat{q}}  
\label{eq:trans_Bernstein}
\end{equation}

The upper and lower bounds on the uncertain function can be found by evaluating ${Y}^b_{\text{max}} =  \text{max}(\boldsymbol{y}^b)$ and ${Y}^b_{\text{min}} = \text{min}(\boldsymbol{y}^b)$, respectively, owing to the range-enclosing property of the Bernstein polynomials, which states that
\begin{equation}
     \text{max}(f(\eta)) \leq {Y}^b_{\text{max}} \hspace{0.5cm} , \hspace{0.5cm} 
     \text{min}(f(\eta)) \geq {Y}^b_{\text{min}}
\label{eq:inq_Bern}
\end{equation}

\begin{figure*}[]
     \centering
\includegraphics[width=0.85\textwidth]{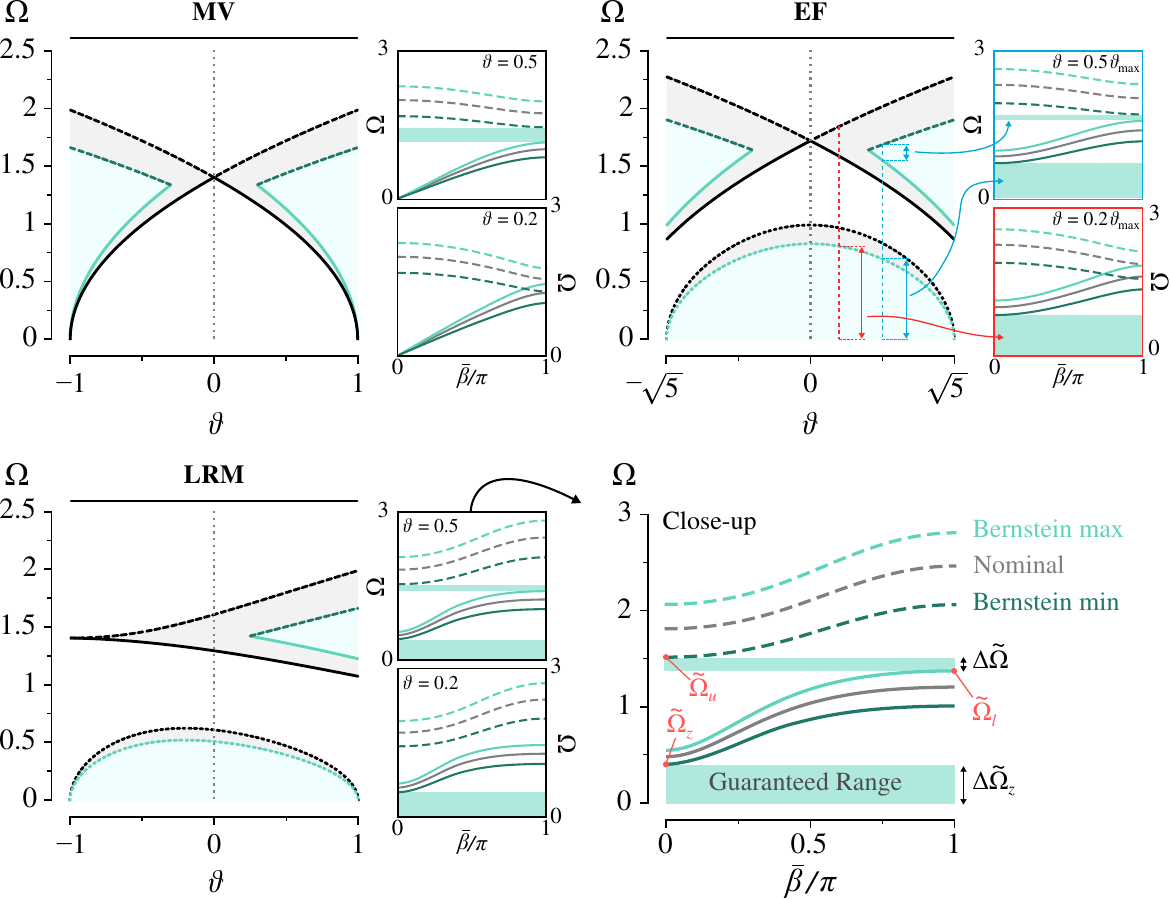}
\caption{Band gap limits as a function of $\vartheta$ for the nominal and uncertain systems with $\tilde{k} = 0.3$ for the MV, EF, and LRM cases. Representative examples of the dispersion diagrams are shown for $\vartheta = 0.5 \vartheta_\text{max}$ and $\vartheta = 0.2\vartheta_\text{max}$. A close-up of the dispersion diagram in the LRM case provides the necessary labels and key to interpret the rest of the dispersion plots}
     \label{fig:Theta_BG_unc}
\end{figure*}

\begin{figure*}[]
\centering
\includegraphics[width=0.8\textwidth]{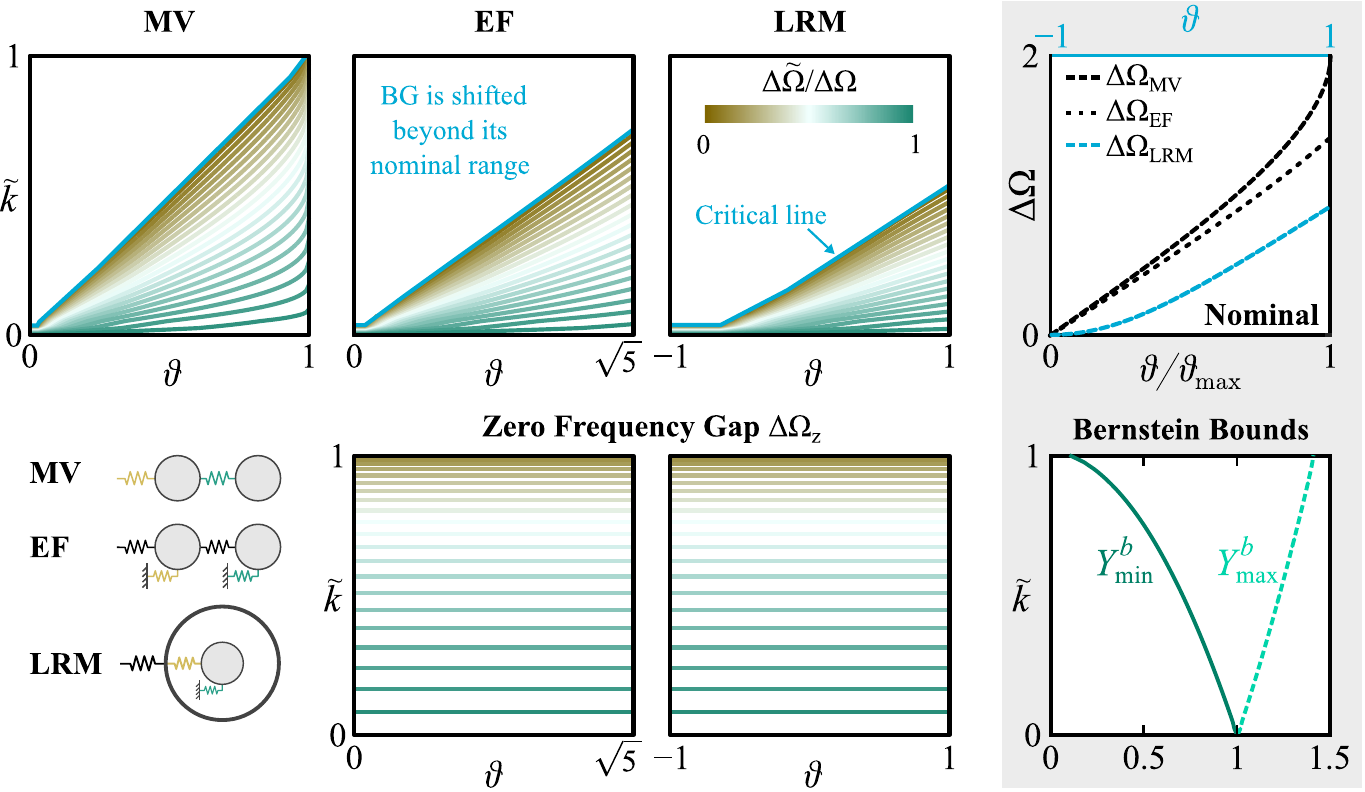}
\caption{Sensitivity of the guaranteed range of the band gap as a function of the control parameter $\vartheta$ and the uncertainty percentage $\tilde{k}$ for the MV, EF, and LRM cases (Plots for the zero frequency band gap are only for the EF and LRM cases). Nominal band gap widths and Bernstein bounds are provided in the rightmost panel for reference}
\label{fig:ktilde_theta}
\end{figure*}

\subsection{Uncertainty in Dispersion Relation}
\label{sec:DispRel_Uncertainty}
Uncertainties in the dispersion properties of elastic metamaterials can arise from uncertainties in the stiffnesses, masses, or both. Without loss of generality, we limit the investigation here to uncertainties in the stiffness $k$. Similar procedure can be followed in the presence of multiple uncertain variables (See \cite{singh2010polynomial} for example). Assuming a uniform distribution across the domain $k_\text{min} \leq k \leq k_\text{max}$ and knowing that $\eta \in [-1,1]$, the parameterization  $k(\eta) = k_{\text{m}} + k_{\text{s}}\eta$ suffices to represent the uncertain variable $k(\eta)$ \cite{singh2010polynomial} 
\begin{subequations}
\begin{equation}
    k_{\text{m}} = \frac{k_\text{max} + k_\text{min}}{2}
\end{equation}
\begin{equation}
    k_{\text{s}} = \frac{k_\text{max} - k_\text{min}}{2}
\end{equation}
\end{subequations}
such that $k_{\text{m}}$ and $k_{\text{s}}$ are the mean and the stiffness variation slope, respectively. Choosing $k_{\text{m}} = k_0$ and $ k_{\text{s}} = k_0 \tilde{k}$ which gives $k(\eta) = k_0(1 + \tilde{k} \eta)$, where $\tilde{k}$ is referred to as the uncertainty percentage and, for the present analysis, spans the range $\tilde{k} \in [0,1]$.  Substituting $k(\eta) = k_0(1 + \tilde{k} \eta)$, the uncertain dispersion relation branches can be expressed as
\begin{equation}
    \tilde{\omega}_\pm(\vartheta,\bar{\beta}) = f(\eta) \hspace{0.1cm} \omega_\pm (\vartheta,\bar{\beta})
\end{equation}
where
\begin{equation}
    f(\eta) = \sqrt{1+ \tilde{k} \eta}
    \label{eq:omega_pm_xi}
\end{equation}

Next, we aim to compute the integrals in Eq.~(\ref{eq:y_q}) with the substitution of Eq.~(\ref{eq:omega_pm_xi}). Since $g(\eta)$ is a constant equal to $\frac{1}{2}$ for a uniform distribution PDF, $g(\eta)$ cancels out in both the numerator and denominator of Eq.~(\ref{eq:y_q}) and is, therefore, excluded from the analysis thereafter. Making use of Eq.~(\ref{eq:Leg_poly}), the numerator integral in Eq.~(\ref{eq:y_q}) now reads
\begin{equation}
    \mathcal{I}_\text{N} =  \frac{1}{2^p}\sum_{q = 0}^{\lfloor \frac{p}{2} \rfloor} (-1)^q {p \choose q} {2(p-q) \choose p} \int_{-1}^1 \eta^{p-2q} \sqrt{1+ \tilde{k} \eta} \ d\eta
    \label{eq:I}
\end{equation}
and the evaluation of the integral in Eq.~(\ref{eq:I}) results in
\begin{widetext}
\begin{equation}
    \mathcal{I}_\text{N} =  \frac{1}{2^p}\sum_{q = 0}^{\lfloor \frac{p}{2} \rfloor} (-1)^q {p \choose q} {2(p-q) \choose p} \left[\frac{2}{3} \frac{(-1)^{2q-p}}{\tilde{k}^{p-2q+1}} (1+\tilde{k} \eta)^{\frac{3}{2}} {}_{2} \mathcal{F}_{1} \left(\frac{3}{2},2q-p;\frac{5}{2}; 1+\tilde{k} \eta \right) \right] \Bigg|_{-1}^1
    \label{eq:I2}
\end{equation}
\end{widetext}

The last term in Eq.~(\ref{eq:I2}), i.e.  ${}_{2} \mathcal{F}_{1}$, is the hypergeometric series and, given that $2q-p$ is a negative integer, is expressed as \cite{xiu2002wiener}
\begin{equation}
    {}_{2} \mathcal{F}_{1} \left(\frac{3}{2},2q-p;\frac{5}{2}; 1+\tilde{k} \eta \right) = \sum_{l = 0}^{p-2q} \frac{\left(\frac{3}{2} \right)_l (2q-p)_l}{\left(\frac{5}{2} \right)_l l!} \left(1+\tilde{k} \eta \right)^l
    \label{eq:F_12}
\end{equation}
where 
\begin{equation}
(\mathcal{C})_l = 
    \begin{cases}
    1& l = 0\\
    \mathcal{C}(\mathcal{C}+1) \dots (\mathcal{C}+l-1)& l = 1,2, \dots\\
    \end{cases}
\end{equation}

In the case of $2q-p = 0$, Eq.~(\ref{eq:F_12}) automatically reduces to unity. Upon substituting the integral limits and performing a few mathematical manipulations, we arrive at
\begin{widetext}
\begin{equation}
    \mathcal{I}_\text{N} = \sum_{q = 0}^{\lfloor \frac{p}{2} \rfloor}   \sum_{l = 0}^{p-2q}  \frac{(-1)^{3q-p} (2q-p)_l}{2^{p-1} \tilde{k}^{p-2q+1} (2l+3) l!} {p \choose q} {2(p-q) \choose p} \Bigg[ (1+\tilde{k})^{l+\frac{3}{2}} - (1-\tilde{k})^{l+\frac{3}{2}} \Bigg] \
\end{equation}
\end{widetext}

The denominator integral $\mathcal{I}_\text{D}$, on the other hand, can be shown to be a function of $p$, i.e. the order of $\mathcal{P}_p(\eta)$, and takes the following form
\begin{equation}
\mathcal{I}_\text{D} =\frac{2}{2p+1}
\end{equation}

Once the coefficients $Y_p = \mathcal{I}_\text{N}/\mathcal{I}_\text{D}$ are computed, the transformation to the Bernstein bases is achieved via Eq.~(\ref{eq:trans_Bernstein}). The range-enclosing property presented by the inequalities in Eq.~(\ref{eq:inq_Bern}) implies that the upper and the lower bounds of the uncertain band structure can be found by simply calculating the coefficients of the Bernstein polynomial expansion of the uncertain function $f(\eta)$. As such, we exploit the Bernstein bounds to extract the variation in the dispersion diagram and, consequently, the resultant band gaps width/limits can then be predicted in the presence of model uncertainties.

Figure~\ref{fig:Theta_BG_unc} shows the band gap limits as a function of $\vartheta$ for the nominal (i.e. black lines as in Figure~\ref{fig:Theta_BG}) and uncertain systems (light and dark green lines) for all three cases, with an uncertainty percentage of $\tilde{k} = 0.3$ applied throughout. The considered order of the polynomial expansion of the uncertain system is chosen as $n_p = 6$. The confined range between the light and dark green lines indicates the \textit{guaranteed} portion of the nominal band gap width. The dispersion diagram of the uncertain EMs with control parameters of $\vartheta = 0.5 \vartheta_{\max}$ and $\vartheta = 0.2 \vartheta_{\max}$ are provided to facilitate the explanation of the effect of uncertainty on the emerging band gap. For the uncertain dispersion relations, the sub-figures represent the enclosing Bernstein range that sandwiches the optical and acoustical dispersion branches of the nominal case (labeled accordingly in the figure). The maximum and minimum Bernstein bounds in each of the dispersion branches are simply found by multiplying the corresponding dispersion branch equation, i.e. $\Omega_\pm$, by ${Y}^b_{\text{max}}$ and ${Y}^b_{\text{min}}$, respectively. The guaranteed portion of the second band gap is found by the points marked $\tilde{\Omega}_u = {Y}^b_{\text{min}} {\Omega}_u $ and $\tilde{\Omega}_l = {Y}^b_{\text{max}} {\Omega}_l$, the difference of which gives the second band gap width $\Delta \tilde{\Omega}$. It is noteworthy that the guaranteed portion is a function of the uncertainty percentage $\tilde{k}$ as well as the control parameter $\vartheta$, as can be inferred from Figure~\ref{fig:Theta_BG_unc}. In the present example, the gain $\vartheta = 0.2 \vartheta_{\max}$ renders $\tilde{\Omega}_l$ larger than $\tilde{\Omega}_u$ which results in a negative $\Delta \tilde{\Omega}$ and shifts the band gap beyond the design requirement. On the other hand, $\vartheta = 0.5 \vartheta_{\max}$ guarantees around $42 \%$, $20 \%$ and $23 \%$ of the desired second gap for the MV, EF and LRM cases, respectively. In contrast to the second band gap, a portion of the first band gap is always guaranteed and is simply found by setting $\Delta \tilde{\Omega} = \tilde{\Omega}_z$, where $\tilde{\Omega}_z = {Y}^b_{\text{min}}{\Omega}_z$. Given a $\tilde{k} = 0.3$ and $\vartheta = 0.5 \vartheta_{\max}$, for instance, $84\%$ of the desired first gap for both the EF and LRM cases can be guaranteed.

To obtain a comprehensive picture of the effects of both of $\tilde{k}$ and $\vartheta$, a contour map can be constructed by sweeping the entire range of the two variables against the percentage of the guaranteed band gap width $\Delta \tilde{\Omega}/\Delta \Omega$. The swept range of $\vartheta$ covers the entire design space for the LRM and only half of that in both the MV and EF cases owing to the symmetry of their band gap limits about $\vartheta = 0$. The results of such simulation is depicted in Figure~\ref{fig:ktilde_theta}. As observed from the figure, the threshold that ensures a portion of the desired band gap is contingent on the type of the metamaterial with the MV case being the least sensitive to the uncertainty.  Another intriguing observation is that the zero frequency band gap is only affected by the uncertainty percentage $\tilde{k}$ and is generally less sensitive to it than the second band gap. Furthermore, the percentage of the guaranteed zero frequency band gap width is precisely equal to the minimum Bernstein bound ${Y}^b_{\text{min}}$. The nominal band gap width and the calculated Bernstein bounds, i.e. ${Y}^b_{\text{max}}$ and ${Y}^b_{\text{min}}$, are provided in the rightmost column of Figure~\ref{fig:ktilde_theta} for reference.

\subsection{Monte Carlo Simulations: Dispersion Relation}

\begin{figure*}[]
     \centering
\includegraphics[width=0.77\textwidth]{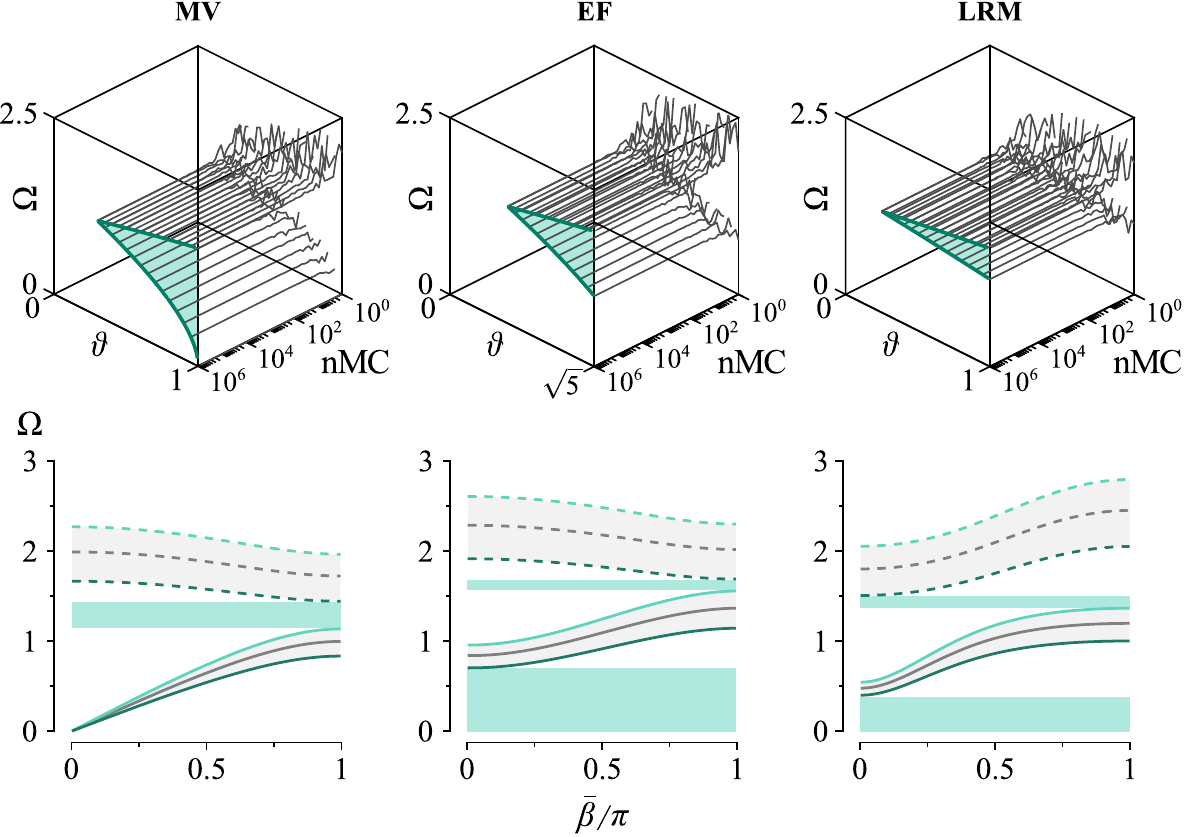}
\caption{\textit{Top:} Monte Carlo (MC) simulations for the non-zero-frequency band gap limits in the presence of uncertainties to verify the predictions of Figure~\ref{fig:Theta_BG_unc}. The latter are indicated with solid green lines for reference. \textit{Bottom:} Monte Carlo simulations for the dispersion diagrams with $\vartheta = 0.5 \vartheta_{\max}$. The grey shaded regions represent the MC simulations}
     \label{fig:MC_disp}
 \end{figure*}

Our analysis thus far has analytically derived the Bernstein bounds on the dispersion relations of uncertain EMs. To verify these predictions, Monte Carlo (MC) simulations are commonly exercised to test the Bernstein bounds against the number of MC samples. As the number of simulation samples increases, we expect the converged bounds based on the MC simulation to coincide with the analytical counterpart, both for the individual dispersion diagrams as well as the band gap limits as a function of $\vartheta$, i.e. $\tilde{\Omega}_l$ and $\tilde{\Omega}_u$. The MC process is executed by generating an EM lattice with the uncertain variable $k$ randomly sampled from its uncertain range, i.e., $k_{\text{min}}\leq k \leq k_{\text{max}}$. The top panel of Figure~\ref{fig:MC_disp} shows MC simulations for the band gap limits of the uncertain EMs considered earlier in Figure~\ref{fig:Theta_BG_unc}. Corresponding dispersion relations at $\vartheta = 0.5 \vartheta_{\text{max}}$ are provided in the bottom panel. As the number of runs increases, Bernstein bounds numerically obtained via MC converge to their analytical counterparts (indicated with green lines for convenience) for all three cases; thus providing confidence in the stability of the MC estimates. For the dispersion diagram simulations, $k$ is sampled 10,000 times and all the dispersion branches lie within the grey area covering the range of the Bernstein bounds as expected.
 


\begin{figure*}[]
     \centering
\includegraphics[width=0.9\textwidth]{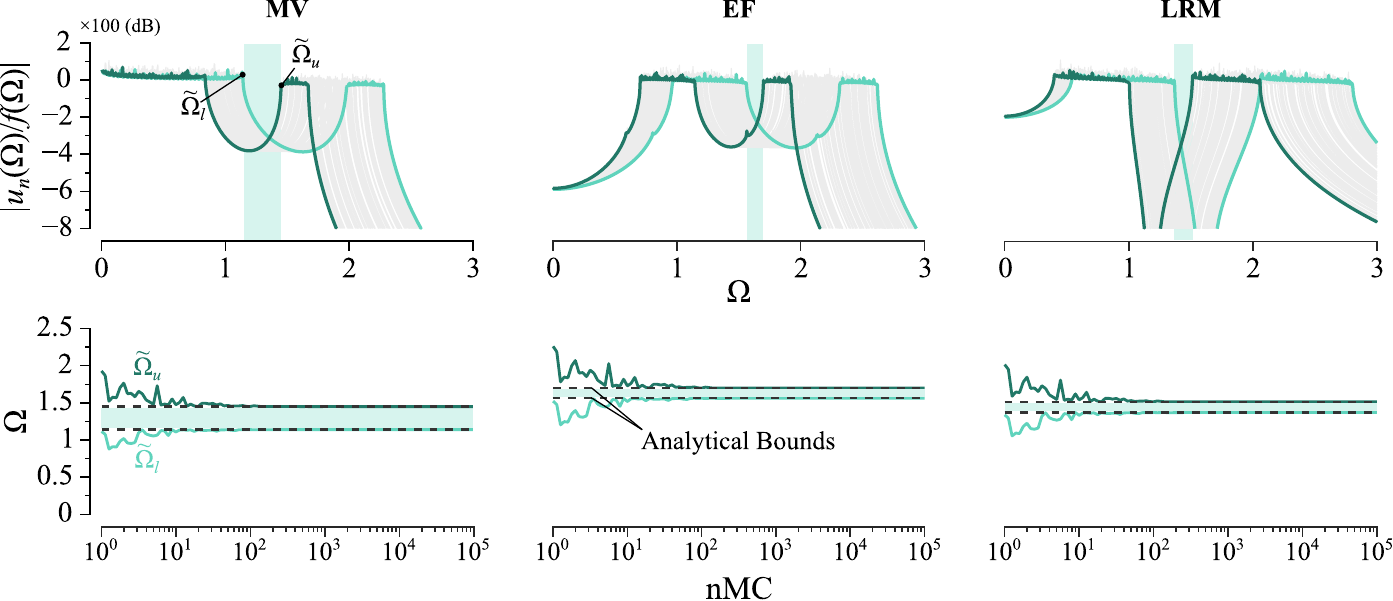}
     \caption{\textit{Top:} Monte Carlo simulations for the frequency response function (FRF) of the finite EMs with $\vartheta = 0.5 \vartheta_{\text{max}}$, $n = 40$ and $\text{nMC} = 10^2$, for the cases where $k = k_{\text{min}}$ (dark lines) and $k = k_{\text{max}}$ (light lines). \textit{Bottom:} Convergence of MC bounds with the increasing number of MC runs for the natural frequencies flanking the guaranteed band gap, i.e. $\tilde{\Omega}_u$ and $\tilde{\Omega}_l$, serving as numerical validation of the bounds predicted in Figure~\ref{fig:Theta_BG_unc}}
     \label{fig:MC_FRF}
\end{figure*}

\subsection{Monte Carlo Simulations: Finite EMs}

The uncertainty quantification is further verified via MC simulations of the finite realizations of the three EM types based on the FRFs derived earlier. In Figure~\ref{fig:MC_FRF}, random values of the uncertain stiffness range are sampled to simulate the FRF, shown here for $\vartheta = 0.5 \vartheta_{\text{max}}$, a number of cells $n = 40$ and $\text{nMC} = 10^2$. A small sample of the MC runs are shown here for the clarity of presentation. FRFs corresponding to MC samples are plotted as grey lines and the analytical bounds of the Bernstein polynomial are superimposed with light and dark green ones. It can be clearly observed that all the MC runs are sandwiched within the analytical Bernstein bounds. In finite metamaterials, it has been shown that the band gap limits coincide with the two natural frequencies flanking the band gap, the latter being a natural-frequency-free range \cite{albabaa2017PC}. As such, the global set of natural frequencies of the finite metamaterials can be effectively split into two groups representing poles which lie on the acoustical and optical branches, respectively. As a result, the range of the guaranteed band gap in the finite EMs will be assessed based on the highest-frequency pole in the first group and the lowest-frequency pole in the second one. The guaranteed band gap limits (i.e. $\tilde{\Omega}_l$ to $\tilde{\Omega}_u$ as marked on Figure~\ref{fig:MC_FRF}) are then extracted. Finally, an MC convergence test of these poles (natural frequencies) is carried out and the range predicted by the analytical Bernstein bounds is shown to perfectly agree with that of the FRF-based analysis (bottom panel of Figure~\ref{fig:MC_FRF}). 

\begin{figure}[h!]
     \centering
\includegraphics[width=0.33\textwidth]{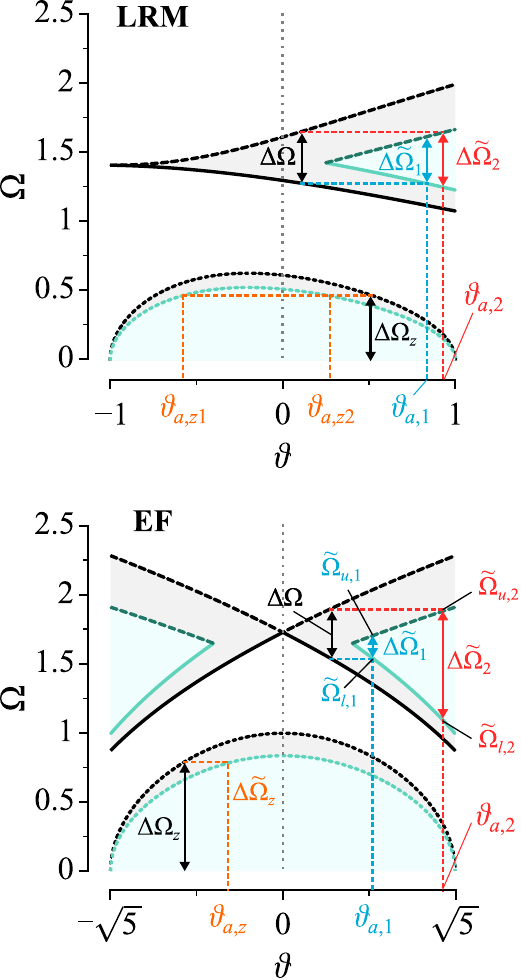}
 \caption{An illustrative schematic of the procedure of Algorithm~\ref{alg:pseudo} used to generate the design maps displayed in Figures~\ref{fig:th_nom_act1},~\ref{fig:th_nom_act2} and~\ref{fig:collective_2D}e}
     \label{fig:algorithm}
\end{figure}

\begin{figure*}[]
     \centering
\includegraphics[width=0.8\textwidth]{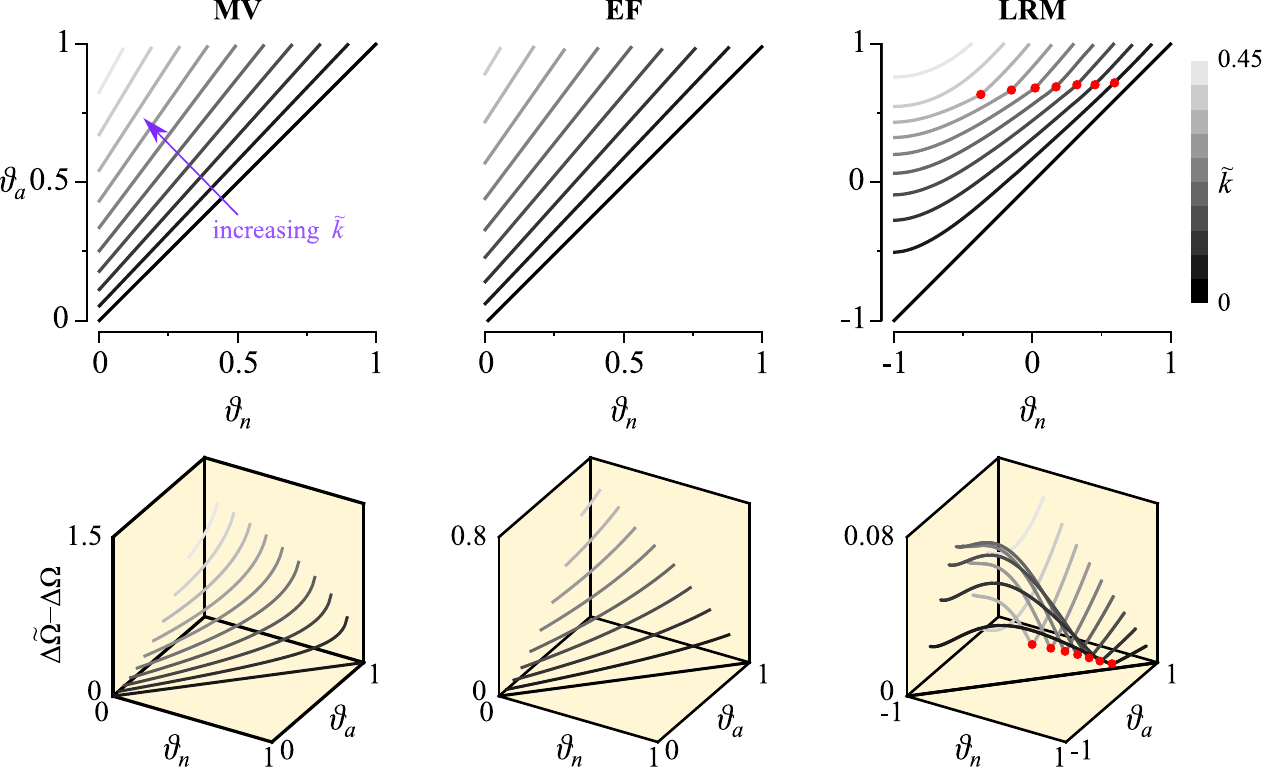}
     \caption{Nominal and recommended gains, $\vartheta_n$ and $\vartheta_a$, respectively, for the MV, EF, and LRM cases. The difference between the nominal band gap and the actual one is plotted in the bottom panel for all combinations of $\vartheta_n$ and $\vartheta_a$. Red circles in the LRM case indicate scenarios where an exact mapping between $\vartheta_n$ and $\vartheta_a$ with $\Delta \tilde{\Omega} - \Delta \Omega = 0$ becomes possible (Actual and nominal gain values are normalized with respect to $\vartheta_\text{\text{max}}$)}
     \label{fig:th_nom_act1}
\end{figure*}

\section{Design guidelines for uncertain EMs}


An uncertain stiffness $k$ yields uncertainty in the band gap width generated by an EM, as has been thoroughly demonstrated. The presence of uncertainty naturally leads to a deviation in the actual metamaterial response from the desired performance based on nominal system parameters. Thus, the goal of this section is to propound a guideline for the designer to determine a value of $\vartheta$ that best meets the design requirements for an uncertain EM. Building on the analysis established in Section~\ref{sec:uncertain_sec} for a given uncertainty percentage $\tilde{k}$, we propose an algorithm that maps a nominal control parameter to an actual or corrected one, denoted as $\vartheta_n$ and $\vartheta_a$, respectively; both of which are normalized with respect to $\vartheta_{\text{max}}$. Successfully tuning the band structure of a given EM by exciting the system with a corrected control parameter $\vartheta_a$ ensures that the desirable band gap location $\Delta {\Omega}$ (or at least a portion of it) lies within the actual emerging band gap $\Delta \tilde{\Omega}$ with the possibility of $\Delta \tilde{\Omega} \geq \Delta \Omega$. A solution, however, is not always guaranteed and will depend on the prescribed uncertainty percentage $\tilde{k}$. The proposed algorithm is graphically illustrated in Figure~\ref{fig:algorithm} and a pseudo code is provided in Algorithm~\ref{alg:pseudo} below for a specific value of $\tilde{k}$. It is worth reemphasizing that only half the range of $\vartheta$ values needs to be scanned for the MV and EF cases, unlike the LRM where the entire range of $\vartheta$ is swept.

\begin{algorithm}[H]
\caption{Nominal and Actual Control Parameter Search}\label{alg:pseudo}
\begin{algorithmic}[]
\For{$\vartheta_n$ between -1 to 1}\Comment{or from 0 to 1 for MV/EF}
\State Solve $\Omega_l(\vartheta_\text{n}) - \tilde{\Omega}_{l,1}(\vartheta_\text{a}) = 0$
\If{$\Delta \tilde{\Omega}_1(\vartheta_\text{a}) \geq \Delta \Omega(\vartheta_\text{n})$ }

\State Record $\vartheta_a$, $\Delta \tilde{\Omega}_1(\vartheta_\text{a}) - \Delta \Omega(\vartheta_\text{n})$
\Else{}
\State Solve $\Omega_u(\vartheta_\text{n}) - \tilde{\Omega}_{u,2}(\vartheta_\text{a}) = 0$
\If{$\Delta \tilde{\Omega}_2(\vartheta_\text{a}) \geq \Delta \Omega(\vartheta_\text{n})$ }
\State Record $\vartheta_a$, $\Delta \tilde{\Omega}_2(\vartheta_\text{a}) - \Delta \Omega(\vartheta_\text{n})$

\Else{}
\State Continue \Comment{Continue to the next $\vartheta_n$}
\EndIf
\EndIf
\EndFor
\end{algorithmic}
\end{algorithm}

\begin{figure}[h]
     \centering
\includegraphics[width=0.49\textwidth]{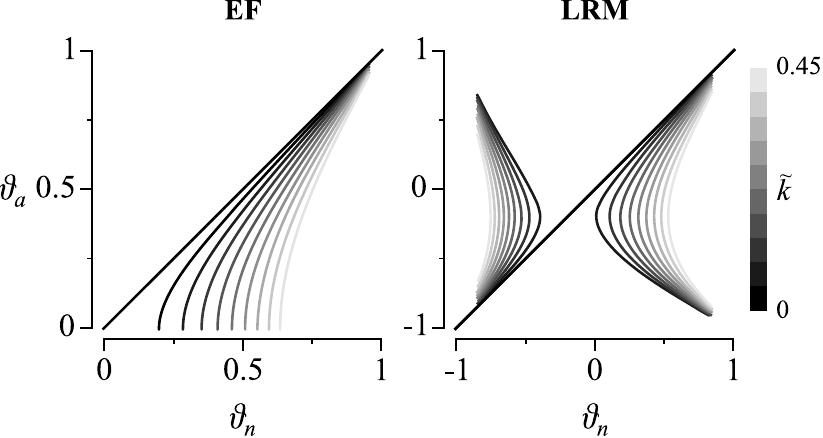}
     \caption{Normalized nominal and recommended gains, $\vartheta_n$ and $\vartheta_a$, respectively, for the first (i.e. zero frequency) band gap in the EF and LRM cases}
     \label{fig:th_nom_act2}
\end{figure}

Based on Algorithm~\ref{alg:pseudo}, a mapping between $\vartheta_n$ and $\vartheta_a$ pertaining to the second band gap can be established for various values of $\tilde{k}$, which is shown in Figure~\ref{fig:th_nom_act1}. As can be seen in the lower panel of Figure~\ref{fig:th_nom_act1}, the actual control parameter $\vartheta_a$ tends to be higher than the nominal $\vartheta_n$ and the actual gap is sequentially larger. The MV case generates a significantly larger actual band gap than the other two types, followed by the EF and then the LRM. Additionally, the actual band gap in the LRM case can be equal to its nominal counterpart, i.e. $\Delta \tilde{\Omega} - \Delta \Omega = 0$. Such occurrences are marked with red circles in Figure~\ref{fig:th_nom_act1}. A similar analysis is carried out for the first (i.e. zero frequency) band gap and the outcome is displayed in Figure~\ref{fig:th_nom_act2}. Interestingly, in this case, an exact matching between the actual and nominal band gap widths is always possible. The LRM case, given its asymmetric band gap limits profile, can exhibit two distinct solutions which match the actual and nominal control parameters, with $\vartheta_a$ possibly being larger or smaller in magnitude relative to the nominal gain. For the EF case, however, $\vartheta_a$ is always smaller than $\vartheta_n$ as depicted in Figure~\ref{fig:algorithm}. 


\section{Extension to 2D Lattices}

\begin{figure*}[]
     \centering
\includegraphics[]{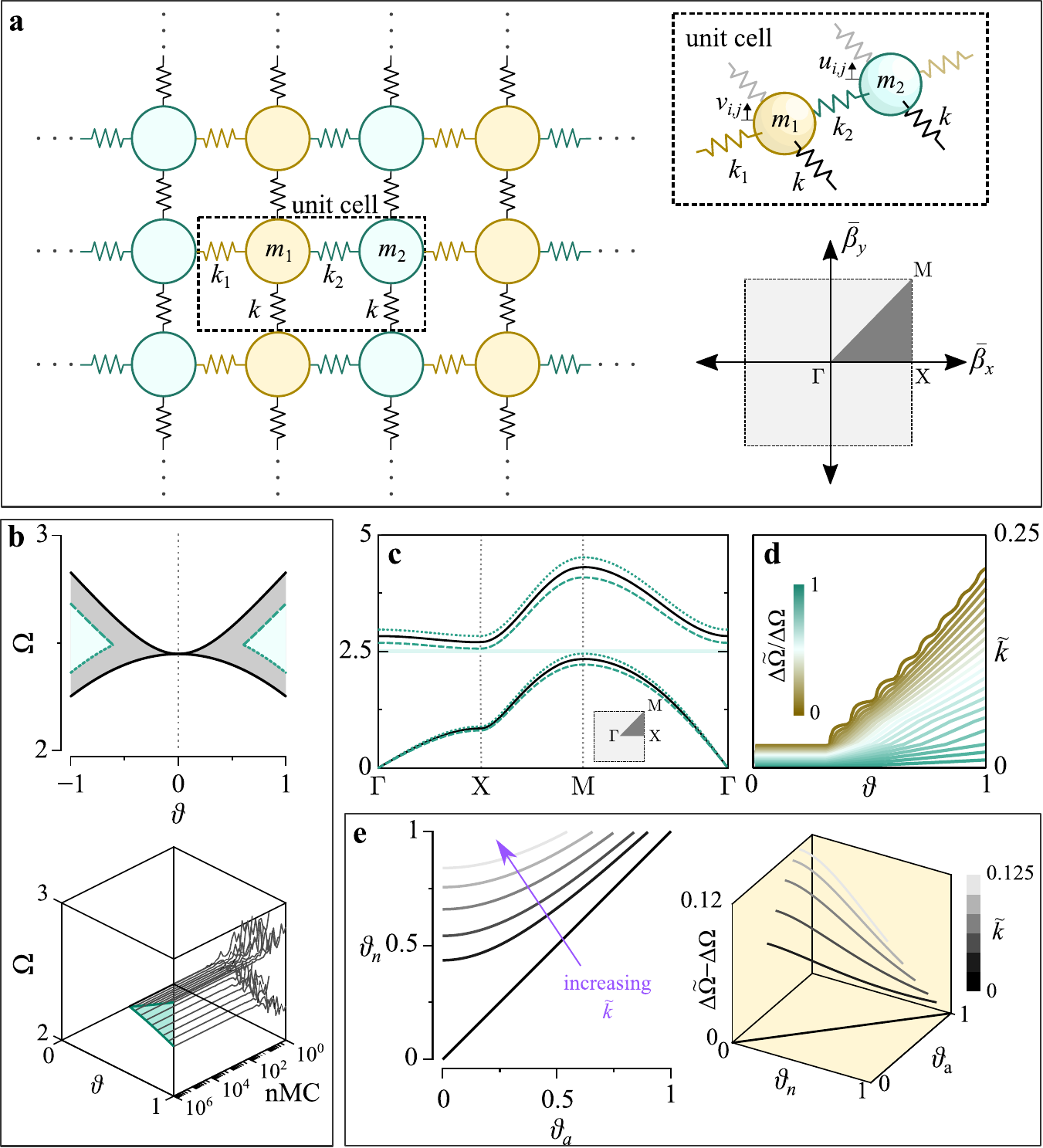}
     \caption{Uncertainty analysis of the 2D square lattice: (a) Schematic diagram of a 2D square phononic crystal lattice undergoing transverse motion and the definition of its self-repeating unit cell. The irreducible Brillouin zone (IBZ) is also shown for reference. (b) Band gap limits for the nominal and uncertain systems with $\tilde{k} = 0.25$ as well as the MC validation. (c) An example of the dispersion relation with $\vartheta = 0.75$ and $\tilde{k} = 0.25$. (d) The percentage of guaranteed band gap as a function of the uncertainty percentage $\tilde{k}$. (e) Design map of the actual and nominal control gains and the associated discrepancy between the actual and nominal band gap widths}
     \label{fig:collective_2D}
 \end{figure*}

The entire framework presented here is not limited to 1D systems and can be conveniently extended to 2D lattices. We briefly present the scenario of a diatomic square lattice, shown in Figure~\ref{fig:collective_2D}a as a case in point. A 2D lattice is characterized by two wavenumbers, $\bar{\beta}_x$ and $\bar{\beta}_y$, pertaining to wave propagation in the $x$ and 
$y$ directions, respectively. Similar to 1D systems, the final form of the dispersion relation is identical to Eq.~(\ref{eq:disp_gen_form}) with the parameters $\alpha_0$, $\alpha_1$ and $\alpha_2$ now being functions of both $\bar{\beta}_x$ and $\bar{\beta}_y$. An in-depth derivation of the dispersion relation of a 2D phononic crystal is provided in Appendix~\ref{app:2D_PC}. Unlike the 1D case, however, a mass ratio of $\mu = \frac{m_2}{m_1}$ of $\frac{1}{3}$ or $3$ is required to initiate a complete band gap with $0<|\vartheta| \leq 1$.

Since the dispersion relation of the 2D lattice takes on an identical form as that of the 1D case, the uncertainty analysis becomes analogous to that of the 1D lattice and the equations presented in Section~\ref{sec:DispRel_Uncertainty} remain intact. The overall results of the uncertainty quantification are laid out in Figure~\ref{fig:collective_2D} for a mass ratio of $\frac{1}{3}$. Figures~\ref{fig:collective_2D}b through f represent the guaranteed band gap limits and the MC validation for $\tilde{k} = 0.1$, the uncertain dispersion diagram at $\vartheta = 0.75$, the percentage of the guaranteed band gap width as a function of $\tilde{k}$ and $\vartheta$, as well as the actual versus nominal control parameters with the associated increase in actual band gap width. The general behavior of the 2D lattice in the presence of uncertainty is similar to its 1D counterpart (e.g. Figures~\ref{fig:th_nom_act1} and \ref{fig:th_nom_act2}). Nonetheless, we observe that the 2D case is more sensitive to uncertainties in the system parameters. For instance, it is possible for the actual band gap to shift beyond the nominal range with $\tilde{k}$ as small as 0.23 at the maximum value of the control parameter $\vartheta_{\max}$.

\section{Conclusions}

This work has presented a robust framework to quantify uncertainty effects on the dispersive wave mechanics as well as the finite structural response of three broad elastic metamaterial (EM) configurations. The scheme, which was adopted based on the Polynomial Chaos theory, accurately predicts the effect of the uncertain stiffness on both the location and width of emerging phononic and local resonance band gaps in one- and two-dimensional EMs. It was shown that band gap may shift beyond its nominal range and the range of variation in the dispersion relation has been quantified using the range-enclosing property of the Bernstein bases, which aid the determination of the guaranteed portion of the nominal band gap width, if any. The results from the dispersion based analysis have been further verified using the end-to-end transfer functions of the finite metamaterials. Further, the predicted Bernstein bounds have been confirmed via a set of standard Monte Carlo simulations. The established analysis and consequent guidelines are invaluable in the design of EMs in the presence of inevitable manufacturing uncertainties and with the crucial need to meet a desired performance.

\section{Acknowledgment}
The authors acknowledge the support of this work from the US National Science Foundation through Award no. 1847254 (CAREER) as well as the NY State Center of Excellence in Materials Informatics.

\newpage

%


\newpage

\appendix
\section{Mass and stiffness matrices for the different EM unit cells}
\label{app:M_and_K}

The mass and stiffness matrices of an LRM are given by

\begin{subequations}
\begin{align}
\mathbf{M}_c= \mathbf{diag}\Big[m, m,0 \Big]
\label{eq:M_c}
\end{align}
\begin{align}
\mathbf{K}_c=
\begin{bmatrix}
k_1 + k_2 & -k_1  & 0 \\
-k_1 & k_1 + k & - k\\
0 & -k & k\\
\end{bmatrix}
\end{align}
\end{subequations}


For the other unit cell types, the mass matrix is the same as Eq.~(\ref{eq:M_c}) with the following stiffness matrices for the material variation (MV) and elastic foundation (EF) cases, respectively

\begin{subequations}
\begin{align}
\mathbf{K}_c=
\begin{bmatrix}
k_1+k_2  & -k_2  & -k_1 \\
-k_2 & k_2 & 0 \\
-k_1 & 0 & k_1\\
\end{bmatrix}
\end{align}

\begin{align}
\mathbf{K}_c=
\begin{bmatrix}
k_1 + 2k  & -k &  - k\\
-k &  k_2 + k & 0 \\
-k & 0 & k \\
\end{bmatrix}
\end{align}
\end{subequations}



\section{Global stiffness matrices for finite EMs}
\label{App:K_matrices}
For the MV and EF cases, the stiffness matrices respectively are
\begin{align}
\underset{2n\times 2n}{\mathrm{\mathbf{K}}} =
\begin{bmatrix}
k_2 & -k_2  & 0 & \dots & \dots & 0  \\
-k_2 & k_1+k_2 & -k_1 & \ddots & & \vdots \\
0 & -k_1  & \ddots &  \ddots & \ddots & \vdots  \\
\vdots & \ddots & \ddots & \ddots & -k_1 & 0 \\
\vdots & & \ddots & -k_1 &  k_1+k_2 & -k_2 \\
0 & \dots& \dots& 0 & -k_2 & k_2 \\
\end{bmatrix} 
\label{eq:K1}
\end{align}

\begin{widetext}
\begin{align}
\underset{2n\times 2n}{\mathrm{\mathbf{K}}} =
\begin{bmatrix}
k+k_1 & -k  & 0 & \dots & \dots & \dots & 0  \\
-k & 2k+k_2 & -k & \ddots & & & \vdots \\
0 & -k  & 2k+k_1 & \ddots & \ddots & &\vdots  \\
\vdots &  \ddots  & \ddots & \ddots & \ddots & \ddots &\vdots  \\
\vdots & & \ddots & \ddots & 2k+k_2 & -k &  0 \\
\vdots & & & \ddots &  -k &  2k+k_1 & -k \\
0 & \dots & \dots& \dots& 0 & -k & k+k_2 \\
\end{bmatrix} 
\label{eq:K2}
\end{align}
\end{widetext}

For the LRM case, on the other hand, the stiffness matrix is given by
\begin{subequations}
\begin{equation}
\underset{2n\times 2n}{\mathrm{\mathbf{K}}} =
    \begin{bmatrix}
    k_1\mathbf{I}_n + k\mathbf{\Psi} & -k_1\mathbf{I}_n \\
    -k_1\mathbf{I}_n & (k_1+k_2) \mathbf{I}_n
    \end{bmatrix}
\end{equation}
\end{subequations}
where
\begin{align}
\underset{n\times n}{\mathrm{\mathbf{\Psi}}} =
\begin{bmatrix}
1 & -1  &  &  &\\
-1 & 2 & \ddots & &\\
& \ddots & \ddots & \ddots &\\
& &  \ddots &  2 &  -1   \\
& & & -1 & 1\\
\end{bmatrix} 
\end{align}

\section{Parametric Study: LRM Dispersion Relation}

\label{app:Para_MM}
\begin{figure*}[]
     \centering
\includegraphics[width=\textwidth]{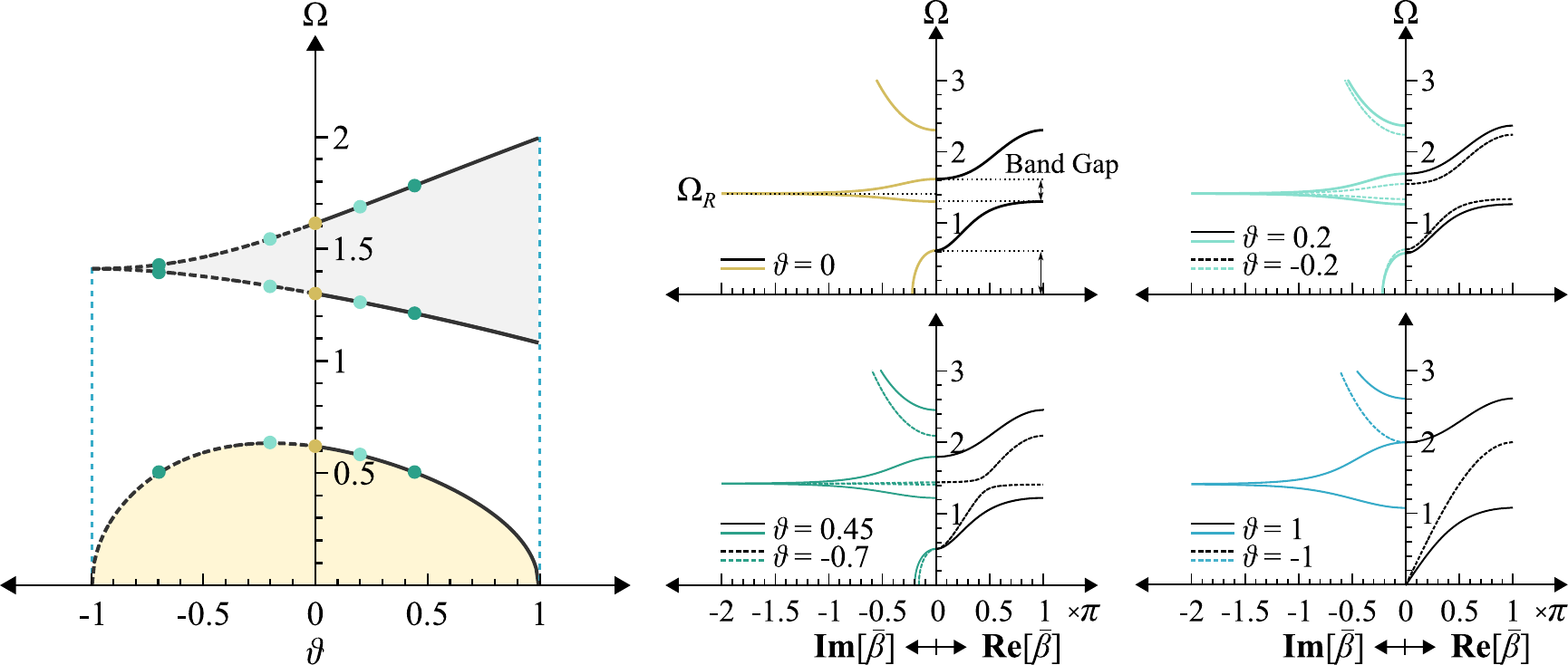}
     \caption{Band Gap limits and dispersion relations of different values of $\vartheta$}
     \label{fig:disp_MM}
\end{figure*}

The non-dimensional dispersion relation of an LRM is derived as
\begin{equation}
    \Omega^4 - (3+\vartheta+4 \sin^2 \frac{\bar{\beta}}{2}) \Omega^2 + (1-\vartheta^2) + 8 \sin^2 \frac{\bar{\beta}}{2} = 0
    \label{eq:disp}
\end{equation}

The dispersion relation can be also written in terms of the dimensionless wavenumber, i.e. $\bar{\beta}(\Omega) = \cos^{-1}(\Phi(\Omega))$, such that
\begin{equation}
    \Phi(\Omega) = 1+ \frac{\Omega^4 - (3+\vartheta)\Omega^2 + (1-\vartheta^2)}{2(2-\Omega^2)}
    \label{eq:phi}
\end{equation}

Note that the anti-resonance associated with this parameterization does not change with the change of $\vartheta$ and remains at $\Omega_R = \sqrt{2}$. This dispersion relation creates two band gaps, one that starts at zero frequency due to the elastic foundation and the other splits the two branches of the dispersion diagram, which happens due to the internal resonator. The maximum attenuation of the zero frequency gap (i.e. first gap) happens at $\Omega = 0$. Substituting back in Eq.~(\ref{eq:phi}), we obtain $\Phi(\Omega) = 1+\frac{1-\vartheta^2}{4}$. This implies that for values of $\pm \vartheta$, the maximum attenuation is identical but does not, however, imply identical band gap widths. If we obtain the same band gap width for two different values of $\vartheta$, then the smaller in magnitude of the two has larger attenuation constants across the entire band gap width.

The configuration under consideration has been shown to have two mass negative regions in literature \cite{pope2010viscoelastic}. The first starts at zero frequency and is attributed to the grounded spring which has an effect similar to an elastic foundation. The evolution of both band gaps as a function of $\vartheta$ can be examined in explicit form using the dispersion relation in Eq.~(\ref{eq:disp}). The limits are found from the solution of the dispersion relation at the limits of the IBZ, i.e. $\bar{\beta} = 0,\pi$, which correspond to $\hat{\beta} = 0,4$, respectively. The two solutions of the dispersion relation $\Omega_\pm(\bar{\beta})$ at the IBZ boundaries, i.e. $\bar{\beta} = 0$ and $\bar{\beta} = \pi$, can be written in the explicit form
\begin{subequations}
\begin{equation}
    \Omega_\pm(\vartheta,0) = \frac{1}{\sqrt{2}} 
    \sqrt{(3+\vartheta) \pm \sqrt{5+6\vartheta+5\vartheta^2}}
\end{equation}
\begin{equation}
    \Omega_\pm(\vartheta,\pi) = \frac{1}{\sqrt{2}} 
    \sqrt{(7+\vartheta) \pm \sqrt{13+14\vartheta+5\vartheta^2}}
\end{equation}
\end{subequations}

The first band gap width is $\Delta \Omega_z = \Omega_-(0)$ while the other is $\Delta \Omega = \Omega_+(0)-\Omega_-(\pi)$. We observe that $\Delta \Omega_z$ reaches its zenith at $\vartheta = -\frac{1}{5}$. This result is mathematically derived from finding the roots of $\frac{\partial \Omega_-(0)}{\partial \vartheta} = 0$, which correspond to $\Delta \Omega_z = 0.6325$. Except for $\vartheta = -\frac{1}{5}$, two values of $\vartheta$ correspond to an identical band gap width 
$\Delta \Omega_z$. These two values can be found analytically by solving the following equation
\begin{equation}
    \vartheta^2 + \Delta \Omega^2_{z} \vartheta - (\Delta \Omega^4_{z}-3\Delta \Omega^2_{z}+1) = 0
\end{equation}
which can be derived from the definition of $\Delta \Omega_z$ by rearranging the equation in terms of $\vartheta$.

\section{Dispersion relation of a 2D lattice}
\label{app:2D_PC}

The dispersion analysis of the 2D lattice shown in Figure~\ref{fig:collective_2D}a is briefly presented here. The stiffness and mass matrices of the unit cell are given by
\begin{subequations}
\begin{align}
\mathbf{M}_c = \text{diag}
\big(m_1,0,m_2,0,0 \big)
\end{align}
\begin{align}
\mathbf{K}_c =
\begin{bmatrix}
k_1+k_2+k & -k & -k_2& -k_1 & 0\\
-k  & k & 0 & 0 & 0\\
-k_2 & 0 & k_2+k & 0 &-k\\
-k_1 & 0 &  0 &  k_1 & 0 \\
0 & 0 & -k & 0 & k\\
\end{bmatrix} 
\end{align}
\end{subequations}
where the degrees of freedom of the unit cell are defined as $\mathbf{x}_c^\text{T} = \{ \mathbf{x}_\text{T} \ \mathbf{x}_\text{B} \ 
\mathbf{x}_\text{RT} \
\mathbf{x}_\text{LT} \
\mathbf{x}_\text{RB} \} = \{ v_{i,j} \ v_{i,j-1} \ u_{i,j} \ u_{i-1,j} \ u_{i,j-1} \}$ (See Eq.~(\ref{eq:gen_dof_2d})). As such, the transformation matrix $\mathbf{Q}$ in Eq.~(\ref{eq:Q_transform}) for this case reduces to
\begin{align}
\mathbf{Q}=
\begin{bmatrix}
1 & 0\\
e^{-\mathbbm{i} \bar{\beta}_y} & 0 \\
0 & 1\\
0 & e^{-\mathbbm{i} \bar{\beta}_x}\\
0 & e^{-\mathbbm{i} \bar{\beta}_y}\\
\end{bmatrix} 
\end{align}
which condenses the degrees of freedom of the system into $\mathbf{x}_c^\text{T}(\omega) = \{\mathbf{x}_\text{T} \ \mathbf{x}_\text{RT} \} = \{ v_{i,j} \ u_{i,j} \} $. Upon deriving the dynamic stiffness matrix in Eq.~(\ref{eq:disp_eq}), the determinant of $\mathbf{D}_c(\bar{\beta}_x,\bar{\beta}_x,\omega)$ is obtained which ultimately yields the dispersion relation in Eq.~(\ref{eq:disp_gen_form}) with the coefficients being $\alpha_2 = 1$ and
\begin{gather}
\alpha_1 = 2 \omega_0^2 \left(1+\frac{1}{\mu} \right) \left(2-\cos(\bar{\beta}_y) \right)\\
\alpha_0 = \frac{2}{\mu} \omega_0^4 \left( 2\left(2-\cos(\bar{\beta}_y) \right)^2 + (\vartheta^2-1) \cos(\bar{\beta}_x) - (\vartheta^2+1) \right)
\end{gather}

Since the main emphasis is placed here on complete band gaps (emerging from a change in the control parameter $\vartheta$), the dynamics of the square lattice require that the masses $m_1$ and $m_2$ be unequal and satisfy the following inequalities: $\mu \leq \frac{1}{3}$ or $\mu \geq 3$, where $\mu=\frac{m_2}{m_1}$. To start with a band gap width of zero at $\vartheta = 0$, the mass ratios $\mu = \frac{1}{3}$ or $3$ are required. This can be proven easily by computing the upper and lower limits of the complete band gap, which are at the X and M edges of the dispersion relation. At X, where $\bar{\beta}_x = \pi$ and $\bar{\beta}_y = 0$, the larger solution of the dispersion relation is the upper limit, while at M with $\bar{\beta}_x = \bar{\beta}_y = \pi$, the smaller solution corresponds to the lower limit. Upon substituting the wavenumber values and computing $\Delta \Omega$, two solutions for the band gap width can be obtained
\begin{widetext}
\begin{equation}
    \Delta \Omega = \sqrt{1+\frac{1}{\mu} + \sqrt{\left(1 - \frac{1}{\mu} \right)^2 + \frac{4}{\mu} \vartheta^2}} - \sqrt{3 \left(1+\frac{1}{\mu}\right) -  \sqrt{9\left(1 - \frac{1}{\mu} \right)^2 + \frac{4}{\mu} \vartheta^2}}
\end{equation}
\begin{equation}
\Delta \Omega =  \sqrt{1+\frac{1}{\mu} - \sqrt{\left(1 - \frac{1}{\mu} \right)^2 + \frac{4}{\mu} \vartheta^2}} - \sqrt{3 \left(1+\frac{1}{\mu}\right) - \sqrt{9\left(1 - \frac{1}{\mu} \right)^2 + \frac{4}{\mu} \vartheta^2}}
\end{equation}
\end{widetext}
for $\mu > 1$ and $\mu<1$, respectively. For $\vartheta =0$, the band gap width reduces to $\Delta \Omega = \sqrt{2} \left(1-\sqrt{3/\mu} \right)$ and $\Delta \Omega = \sqrt{2} \left(\sqrt{1/\mu} - \sqrt{3} \right)$ for $\mu > 1$ and $\mu<1$, respectively. Solving for $\mu$ with $\Delta \Omega = 0$ yields the solutions $\mu = 1/3$ and $\mu = 3$.

\end{document}